\documentclass[12pt,dvips]{article}
\usepackage{rotating}
\usepackage{hhline}
\usepackage{array}
\usepackage{epsfig}

\begin{document}
\tolerance=100000
\thispagestyle{empty}
\setcounter{page}{0}

\newcommand{\nn}{\nonumber}
\newcommand{\be}{\begin{equation}}
\newcommand{\ee}{\end{equation}}
\newcommand{\ba}{\begin{eqnarray}}
\newcommand{\ea}{\end{eqnarray}}
\newcommand{\bann}{\begin{eqnarray*}}
\newcommand{\eann}{\end{eqnarray*}}
\newcommand{\bc}{\begin{center}}
\newcommand{\ec}{\end{center}}
\newcommand{\LEP}{{\sc Lep }}
\newcommand{\LHC}{{\sc Lhc }}
\newcommand{\TESLA}{{\sc Tesla }}
\newcommand{\RRtwo}{\emph{RR2} }
\newcommand{\sts}{\scriptstyle}
\newcommand{\ngs}{\!\!\!\!\!\!}
\newcommand{\rb}[2]{\raisebox{#1}[-#1]{#2}}
\newcommand{\CP}{${\cal CP}$~}
\newcommand{\fg}{Fig.~$\!$}

\newcommand{\sw}{s_{\mbox{\tiny{W}}}}
\newcommand{\cw}{c_{\mbox{\tiny{W}}}}
\newcommand{\MW}{M_{\mathrm{W}}}
\newcommand{\MZ}{M_{\mathrm{Z}}}
\newcommand{\ts}[2]{#1_{\mathrm{#2}}}
\newcommand{\gtot}{g_{\mathrm{tot}}}
\newcommand{\mt}{\tilde{M}}
\newcommand{\smu}{\tilde{\mu}}
\newcommand{\se}{\tilde{e}}
\newcommand{\eR}{e_{\rm R}}
\newcommand{\eL}{e_{\rm L}}
\newcommand{\seR}{\tilde{e}_{\rm R}}
\newcommand{\seL}{\tilde{e}_{\rm L}}
\newcommand{\smR}{\tilde{\mu}_{\rm R}}
\newcommand{\smL}{\tilde{\mu}_{\rm L}}
\newcommand{\slpt}{\tilde{l}}
\newcommand{\schi}{\tilde{\chi}}
\newcommand{\mn}[1]{m_{\tilde{\chi}^0_{#1}}}
\newcommand{\mnsq}[1]{m^2_{\tilde{\chi}^0_{#1}}}

\newcommand{\dmij}{\Delta m^2_{ij}}

\newcommand{\lesim}{\raisebox{-.3ex}{$_{\textstyle <}\atop^{\textstyle\sim}$}}
\newcommand{\gesim}{\raisebox{-.3ex}{$_{\textstyle >}\atop^{\textstyle\sim}$}}
\newcommand{\nslash}{\not{\!n}}
\newcommand{\slsh}[1]{/ \!\!\!\! #1}
\newcommand{\anc}{\rule{0mm}{0mm}}

\begin{titlepage}

\begin{flushright}
DESY 01-066\\
June 2001
\end{flushright}

\vspace{1.2cm}

\begin{center}
{\Large \bf Pair Production of Smuons and Selectrons\\
Near Threshold in $e^+e^-$ and $e^-e^-$ Collisions}\\[1cm]
{\large A.~Freitas, D.J.~Miller and P.M.~Zerwas\\[1cm]
{\it Deutsches Elektronen--Synchrotron DESY, D--22603 Hamburg, Germany}}
\end{center}

\vspace{3cm}

\begin{abstract}
\noindent Non-zero width and Coulomb rescattering effects are analyzed for the
pair production of  smuons and selectrons near the thresholds in $e^+e^-$ and
$e^-e^-$ collisions, respectively.
The excitation curves  are predicted in a gauge-invariant form.
Energy cuts are  designed to reduce irreducible supersymmetric backgrounds.
\end{abstract}

\end{titlepage}

\section{Introduction}

In supersymmetric theories~\cite{Wess:1974kz}, scalar leptons are of particular
interest. Their properties can be determined in $e^\pm e^-$ experiments with
very high experimental precision~\cite{tesla,Martyn}, which provides the
platform for reconstructing the fundamental supersymmetric theory at energies
up to near the Planck scale~\cite{Blair:2001gy}.

The supersymmetric partners of the left- and right-chiral leptons are 
the scalar leptons, $\tilde{l}_{\rm L}$ and $\tilde{l}_{\rm R}$.
Neglecting mixing effects in the first and second
family\footnote{Mixing effects in the smuon and selectron sectors are of the
  order $10^{-5}$ or less
  and can be safely neglected, provided the left- and
  right-chiral sleptons are well separated in mass.}, the chiral states
are identified with the mass eigenstates.

\underline{Scalar muons} (likewise staus) are produced pairwise in
$e^+e^-$ annihilation via
s-channel photon and $Z$ boson exchange~\cite{SLprod}:
\be e^+e^- \to \smu^+_i\smu^-_i \quad (i=\textrm{L/R}) \ee  The cross-sections
rise near the threshold $\sim \beta^3$ as a P-wave process
[$\beta=(1-4m^2_{\smu_i}/s)^{\frac{1}{2}}$ denoting the smuon velocity],
while becoming scale invariant
for large centre-of-mass energies $\sqrt{s}$. In detail, the
Born cross-sections for the production of on-shell smuons are given by
\ba
\sigma [e^+e^- \to \tilde{\mu}^+_i \tilde{\mu}^-_i] 
=& \displaystyle \frac{\pi\alpha^2}{3s} \, \beta^3
 \Bigg[ 1 \!\!\!\!&- \,g_i \, \frac{1-4\sw^2}{\sw\cw} \, \frac{s}{s-\MZ^2}  \nn \\ 
&&+ \,g_i^2 \, \frac{1+(1-4\sw^2)^2}{16\sw^2\cw^2} \left(\frac{s}{s-\MZ^2} \right)^2 \Bigg]
\ea
The masses are denoted by $m_{\smu_i} \; (i =\mathrm{L/R})$ for the
partners of the left- and right-chiral muon states; $\sw$ and $\cw$
are the sine and cosine of the electroweak mixing angle, and the couplings
$g_{\mathrm{L}}= (-1+2\sw^2)/4\sw\cw$ and
$g_{\mathrm{R}}= {\sw}/{\cw}$ denote the $Z$ charges of the L/R
smuons. The fine structure constant $\alpha$ is evaluated at the scale
$\sqrt{s}$.

\underline{Scalar electrons}, on the other hand, are in general produced by
s-channel $\gamma, Z$ and/or t-channel neutralino $\schi_j^0 \; (j=1,...,4)$
exchanges \cite{SEprod}.
In particular, for helicities L and R in $e^+e^-$ and
$e^-e^-$ collisions, the mediating reactions and the types of orbital wave
function near threshold are given by:
\goodbreak
\ba
\eL^+ \eR^- \,/\, \eR^+ \eL^- &\to& \seL^+ \seL^- \,,\, \seR^+ \seR^-
  \quad [\gamma, Z; \, \schi^0] \qquad \mbox{P-wave} \\
\eL^+ \eL^- \,/\, \eR^+ \eR^- &\to& \seR^+ \seL^- \,/\, \seL^+ \seR^-
  \quad [\schi^0]\phantom{\gamma, Z, \, \anc} \qquad \mbox{S-wave} \\[2mm]
\eL^- \eR^- \,/\, \eR^- \eL^- &\to& \seL^- \seR^-
  \phantom{\anc \,/\, \seL^- \seR^-}
  \quad [\schi^0]\phantom{\gamma, Z, \, \anc} \qquad \mbox{P-wave} \\
\eL^- \eL^- \,/\, \eR^- \eR^- &\to& \seL^- \seL^- \,/\, \seR^- \seR^-
  \quad [\schi^0]\phantom{\gamma, Z, \, \anc} \qquad \mbox{S-wave}
\ea

The steep rise $\propto \beta$ render S-wave production processes especially
suitable for precision measurements in threshold scans. Left- and right-chiral
selectrons can be generated in diagonal $\seL\seL, \seR\seR$ and mixed pairs
$\seL\seR$. Since the amplitudes are built up solely by neutralino exchange, the
total cross-sections for $e^-e^-$ collisions can be cast into a simple form:
\be
\sigma [e^-_i e^-_i \to \tilde{e}^-_i \tilde{e}^-_i] 
= \displaystyle \frac{16\pi\alpha^2}{s} \sum_{j=1}^4 \sum_{k=1}^4
  X_{ij}^{2} \, X_{ik}^{*2} \;
  [ G^{jk} + H^{jk} ] \qquad (i =\mathrm{L/R})
\ee
with
\ba
G^{jk} &=& \frac{2}{s} \; \frac{\mn{j}\mn{k}}{(\Delta_j + \Delta_k)}
	\left[ \ln \frac{\Delta_j+\beta}{\Delta_j-\beta} +
	 \ln \frac{\Delta_k+\beta}{\Delta_k-\beta} \right], \nn \\[1ex]
H^{jk} &=& \left\{ \begin{array}{ll}
  \displaystyle \frac{4\beta}{s} \; \frac{\mnsq{j}}{\Delta_j^2 - \beta^2} & j = k
  	\\[3ex]
  \displaystyle 
     \frac{2}{s} \; \frac{\mn{j}\mn{k}}{(\Delta_j - \Delta_k)}
	\left[ \ln \frac{\Delta_k+\beta}{\Delta_k-\beta} -
	 \ln \frac{\Delta_j+\beta}{\Delta_j-\beta} \right] & j \neq k
  \end{array} \right. \label{defsel}
\ea
where
$\Delta_j = 2 (m_{\se_i}^2-\mnsq{j}) / s -1$ and
$X_{ij}=(N^\prime_{j1} - g_i N^\prime_{j2}) / \sqrt{2}.$
The four neutralino masses in the
minimal supersymmetric extension of the Standard Model (MSSM) are
denoted by $m_{\schi_j^0} \; (j=1,...,4)$. The neutralino
mixing matrix $N^\prime$ relates the mass eigenstates with the
eigenstates in the photino-Zino-Higgsino basis.

For the mixed case in $e^+e^-$ collisions:
\be
\sigma [e^+_{\rm L}e^-_{\rm L} \to \tilde{e}^+_{\rm R} \tilde{e}^-_{\rm L}] 
= \displaystyle \frac{16\pi\alpha^2}{s} \sum_{j=1}^4 \sum_{k=1}^4
  X_{{\rm L}j} \, X_{{\rm R}j}^* \, X_{{\rm R}k} \, X_{{\rm L}k}^* \; H^{jk}
\ee
where the definitions of eq.~$\!$(\ref{defsel}) are understood, but with $\Delta_j$
replaced by \mbox{$\overline{\Delta}_j = (m_{\se_{\rm R}}^2 +
m_{\se_{\rm L}}^2 - 2\mnsq{j}) \, / s -1$} and $\beta =
\sqrt{(s-m_{\se_{\rm R}}^2-m_{\se_{\rm L}}^2)^2
 -4 m_{\se_{\rm R}}^2 m_{\se_{\rm L}}^2} \, / s$.
In contrast to the s-channel exchange process, these t-channel exchange
cross-sections approach non-zero asymptotic values for high energies.

\underline{Decays:}
The right-chiral sleptons $\smR$ and $\seR$ decay predominantly into
neutralinos while additional chargino decays are also important for
left-chiral sleptons $\smL$ and $\seL$~\cite{SLdec}:
\ba
\Gamma (\tilde{l}^-_i \to l^-\tilde{\chi}^0_j) &=&
  \alpha |X_{ij}|^2 \, m_{\slpt_i} \left( 1-
  \frac{m_{\schi_j^0}^2}{m_{\slpt_i}^2} \right)^2 \qquad (i =\mathrm{L/R})
  \label{decneu} \\
\Gamma (\tilde{l}^-_{\mathrm{L}} \to \nu_l \tilde{\chi}^-_k) &=&
  \frac{\alpha}{4} |U_{k1}|^2 m_{\slpt_{\mathrm{L}}}
  \left( 1- \frac{m_{\schi_k^-}^2}{m_{\slpt_{\mathrm{L}}}^2} \right)^2 \label{deccha}
\ea
While the notation for the neutralino sector has been defined above,
$m_{\schi_k^{\pm}} \; (k=1,2)$ denote the two chargino masses and $U$ is the
mixing matrix for the negative charginos. For a
right-chiral slepton mass in the range of about $200$ GeV, the widths
are of the order $\Gamma_{\slpt_{\mathrm{R}}} \sim 700$ MeV, while left-chiral
sleptons have widths of order 1 GeV.
  
\section{Smuon Excitation}

\subsection{Threshold Behaviour}

The measurement of the smuon excitation curve near the
threshold is the most accurate method of determining their masses.
Detailed simulations have demonstrated that the right-chiral smuon
mass can be measured at \TESLA to an accuracy of less than $100$~MeV
in this way~\cite{Martyn}, {\it i.e.} to a fraction of a per-mille.  This
experimental error is significantly smaller than the width of the
state.  Therefore, in this report we examine the effect of non-zero
smuon widths on the excitation curves. Additionally, an accurate
theoretical prediction requires the calculation of higher order
effects, notably Coulomb rescattering among the final state
particles. The problems arising in this context are quite similar to
the production of $W$ pairs in $e^+e^-$
annihilations~\cite{higherorder}. For
the sake of clarity we will restrict ourselves to the analysis of
right-chiral smuons\footnote{Because of the low cross-section
it is not possible to reach a precision for left-chiral smuons
comparable to the case for right-chiral smuons.}.

\begin{figure}[tb]
\begin{tabular}{p{7cm}p{7cm}}
 (a) \underline{Double resonance diagram} \vspace{0.3cm}\newline
\phantom{(a} \epsfig{file=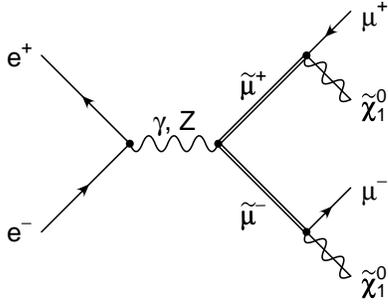,width=6cm} &
 (c) \underline{Coulomb rescattering} \vspace{0.3cm}\newline
\phantom{(c} \epsfig{file=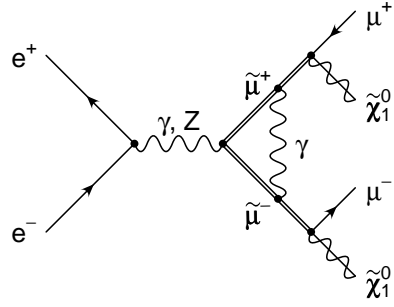,width=6cm} \\
 (b) \underline{Single resonance diagrams} \vspace{0.3cm}\\
\rule{0mm}{0mm}\phantom{(b} \epsfig{file=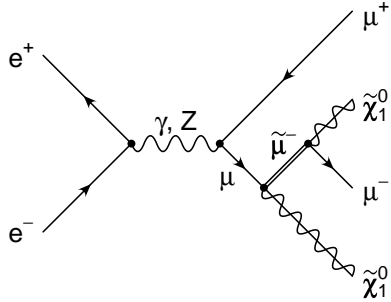,width=6cm} &
\rule{0mm}{0mm}\phantom{(b} \epsfig{file=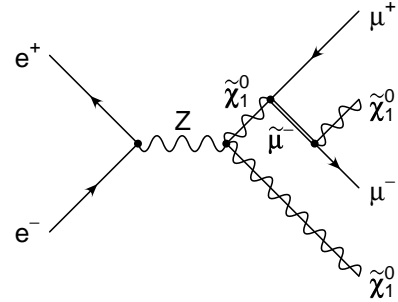,width=6cm}
\end{tabular}
\caption{\it The doubly (a) and singly (b) resonant smuon contributions 
to the process \mbox{$e^+e^- \to \mu^+\mu^- \schi_1^0 \schi_1^0$}; the Coulomb
correction is described by diagram (c).}
\label{fig:signal}
\end{figure}

Decays of the right-chiral smuons to the LSP $\tilde{\chi}_1^0$,
\be \tilde{\mu}_{\rm R}^- \to \mu^- \tilde{\chi}_1^0 \ee
are by far the dominant
decay mode in the MSSM. 
The parameters, masses and couplings, we have
adopted for illustrative examples, correspond to the large $\tan
\beta$ reference point \RRtwo in the \TESLA study Ref.~$\!$\cite{RR}.
The relevant physical parameters are summarized in the appendix.
The pair production of off-shell smuons is described in
this final state by the diagram \fg\ref{fig:signal}(a): 
\be e^+e^- \to \mu^+ \mu^- \tilde{\chi}_1^0 \tilde{\chi}_1^0\ee 
However, the double
resonance diagram must be supplemented by the single resonance
diagrams of \fg\ref{fig:signal}(b) to generate a gauge invariant set
for the off-shell production amplitude.  

If the $[\mu\tilde{\chi}_1^0]$ final states are studied near the
$\smu$ mass, the real $\smu$ propagators must be replaced by the
Breit-Wigner form which explicitly includes the non-zero width of the
resonance state. Near the threshold the substitution leaves the set of
diagrams \fg\ref{fig:signal} gauge invariant if the complex mass
\be m_{\smu}^2 \;\to\; M_{\smu}^2 = m_{\smu}^2 - im_{\smu}\Gamma_{\smu} \ee is
introduced with fixed width $\Gamma_{\smu}$. Applying the substitution
to the propagators in the double and single resonance amplitudes
provides a consistent scheme for smuon production near the threshold.
Even though taking only the doubly resonant contribution
\fg\ref{fig:signal}(a) with fixed width in the covariant gauge (a
scheme generally adopted for experimental simulations~\cite{Martyn})
is of sufficient accuracy at high energies, it fails for
high-precision analyses near the threshold. Indeed the error induced
by this method, about 80 MeV, is generally of the same size as the
error of the experimental measurement.

\begin{figure}[t]
\phantom{(a} \epsfig{file=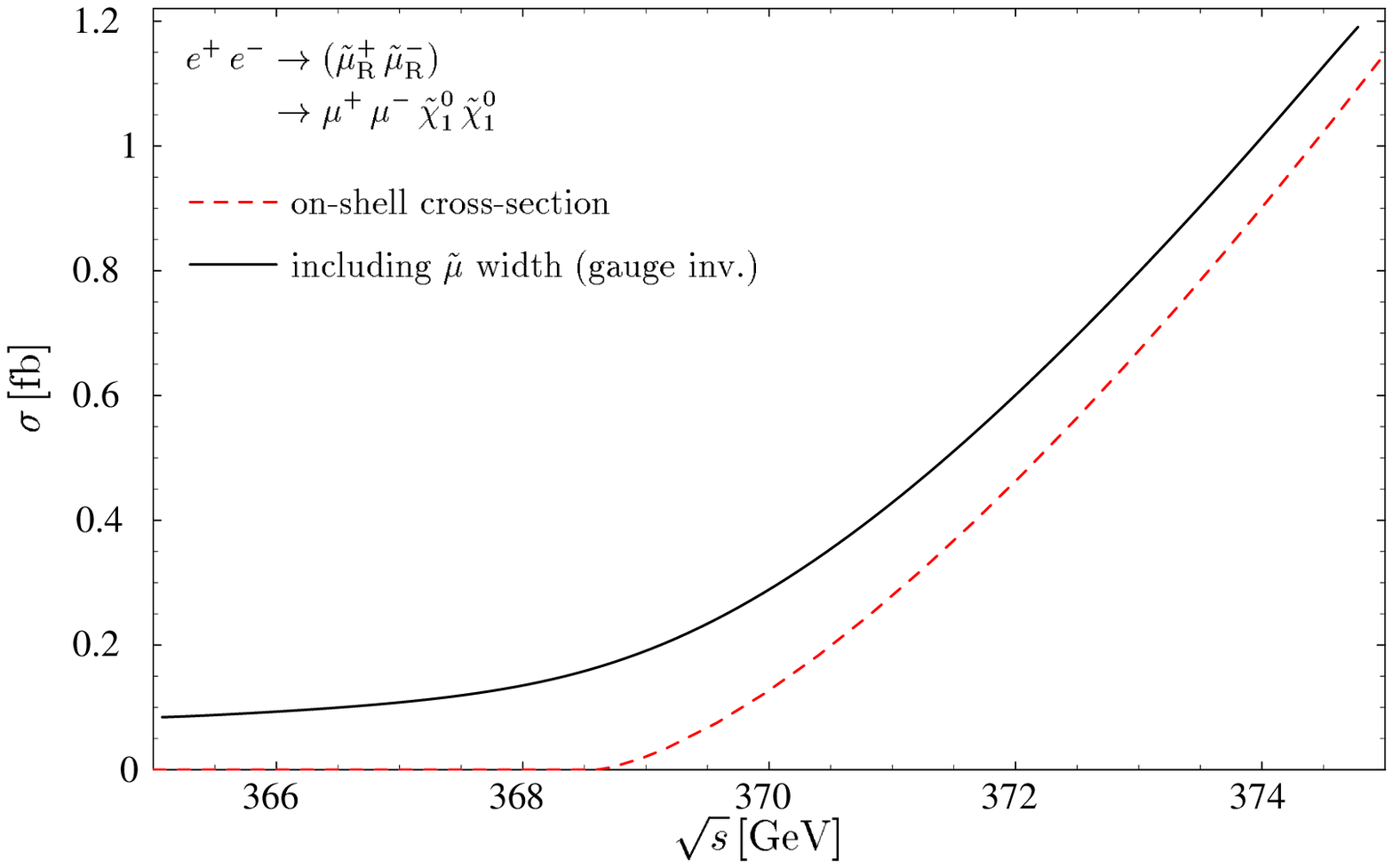,width=5.7in}
\caption{\it The effect of finite $\smu$ width on the $e^+e^- \to \smu_{\rm R}^+ \smu_{\rm R}^- \to \mu^+\mu^-\schi_1^0\schi_1^0$ cross-section.}
\label{fig:sig_at_thresh}
\end{figure}

The effect of non-zero width $\Gamma_{\smu_{\mathrm{R}}}$ on the
$\smu_{\mathrm{R}}^+ \smu_{\mathrm{R}}^-$ excitation curve near the
threshold is demonstrated in \fg\ref{fig:sig_at_thresh} for the \RRtwo
parameters.
While the excitation curve is shown for zero width by the dotted line, 
the prediction of the
excitation curve including the non-zero width effect in a
gauge-invariant form is shown by the full line which can be interpreted 
as the ``signal cross-section''.

The Coulomb interaction due to the photon exchange, cf.
\fg\ref{fig:signal}(c), between slowly moving charged particles gives rise to
large corrections of the threshold cross-section. Independent of spin and
angular momentum, the threshold cross-section for stable particles is
universally modified by $\sigma_{\rm Born} \to (\alpha\pi / 2 \beta) \,
\sigma_{\rm Born}$ at leading order. This Sommerfeld rescattering correction
\cite{Sommf-Sakh} removes one power in the velocity $\beta$ of the threshold
suppression $\sim \beta^{2l+1}$ for $l$-wave production. Multiple photon
exchange further increases the cross-section
but only
by a very small amount.

\begin{figure}[t]
\phantom{(a} \epsfig{file=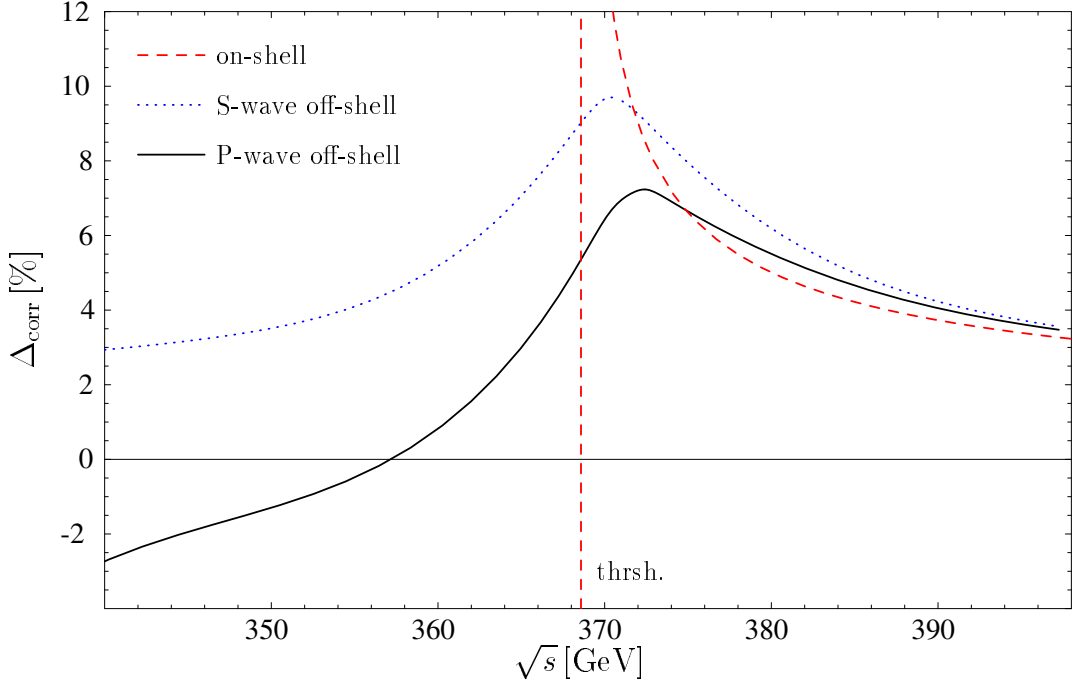,width=5.7in}
\caption{\it
Correction $\Delta_{\rm corr}$ due to Coulomb
rescattering effects relative to the Born cross-section for smuon production.
The realistic case of off-shell P-wave production is compared with the
correction factors for on-shell and off-shell S-wave production.
}
\label{fig:coul}
\end{figure}

However, for the production of off-shell particles the Coulomb singularity is
partially screened \cite{FKoffs}. Moreover, the correction depends on the
orbital angular momentum $l$ in this 
case\footnote{Since the production amplitude
rises proportional to $\beta^l$ near threshold, the maximum contribution is
generated by the configuration of minimum orbital angular momentum $l$. For a
general
vertex $Y \to 2 X$, $l$ is the difference between $2j_X$ and $j_Y$ for
$2j_X < j_Y$, and $0$/$1$ for even/odd $j_Y$ otherwise.}.
For smuon P-wave production one finds at leading order
\ba
\sigma_{\rm Born} &\to& \sigma_{\rm Born} \frac{\alpha\pi}{2\beta_\pm}
\left[ 1- \frac{2}{\pi} \arctan
\frac{|\beta_M|^2 - \beta_\pm^2}{2 \beta_\pm \;
        \Im m \, \beta_M} \right] \Re e \, {\mathcal C}_1 \label{offs} \\
{\mathcal C}_l &=& \left[ \frac{\beta_\pm^2 + \beta_M^2}{2 \beta_\pm^2} \right]^l
 \label{angfac}
\ea
with the generalized velocities
\ba
\beta_\pm &=& \textstyle \lambda^{1/2}(s, m_+^2, m_-^2) \, \big/ s \equiv
  \sqrt{(s-m_+^2-m_-^2)^2-4 m_+^2 m_-^2} \, \Big/ s \\
\beta_M &=& \sqrt{1- 4 M^2/s}
\ea
for the (complex) smuon pole mass $M$
and the smuon virtualities $m_+$
and $m_-$. The off-shellness of the final state particles damps the Coulomb
singularity. 
The damping for S-waves, described by the term in square brackets in
eq.~$\!$(\ref{offs}), is even stronger for waves of higher angular momentum $l$, due
to the additional coefficient ${\mathcal C}_l$, as evident from
\fg\ref{fig:coul}.
In contrast to $W$ pair production, for larger energies $\sqrt{s} - 2 m_{\smu} >
\Gamma_{\smu}$ the off-shell Coulomb correction is slightly enhanced relative
to the on-shell case, which is because the $\beta^3$ dependence of the
production vertex distorts the integration over the Breit-Wigner resonances.

\subsection{SUSY Backgrounds to SUSY Signals}

The general Standard Model backgrounds contributing to the smuon signal
in the process 
$e^+e^- \to \mu^+\mu^- + \slsh{E}$ are
analyzed in Ref.~$\!$\cite{Martyn}.
Most significant are the production of $W$ pairs with the leptonic decay $W \to
\mu \nu$, and of $\gamma/Z$ pairs with the decays $\gamma/Z \to \mu\mu$ and
$\gamma/Z \to \nu\nu$.
They can be reduced by observing that the decay leptons of the $W$'s lie
approximately in an
azimuthal plane and the invariant mass of the lepton pair from $\gamma/Z$
concentrate at the collinear and $Z$ pole, respectively. The appropriate
cuts and the detector acceptance allow a resulting signal efficiency of 50\%.

\begin{figure}[!p]
\flushleft{(a)} \\
\centering\epsfig{file=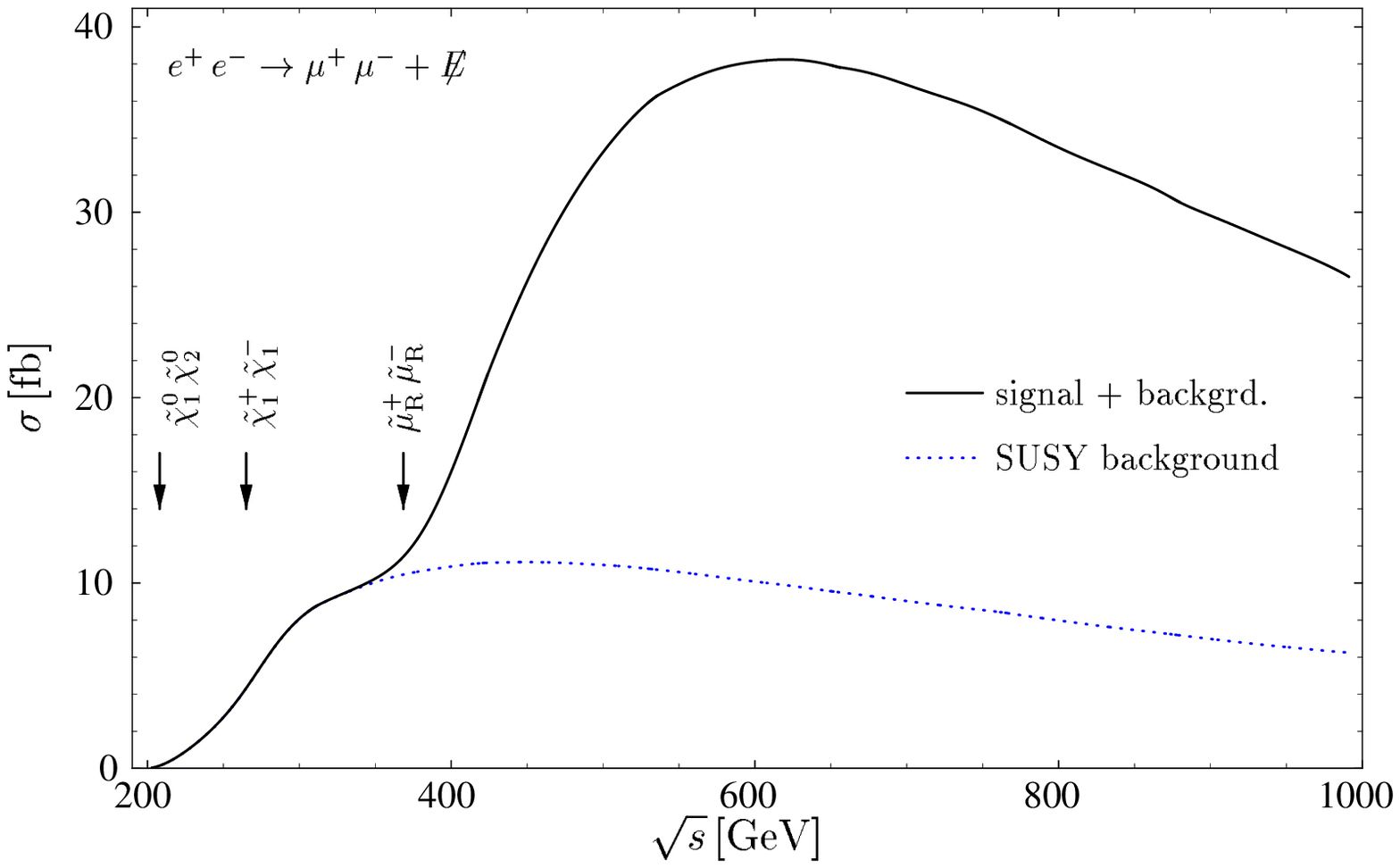,width=5.7in} \\
\flushleft{(b)} \\
\begin{center}
\hspace{0.2cm}
\epsfig{file=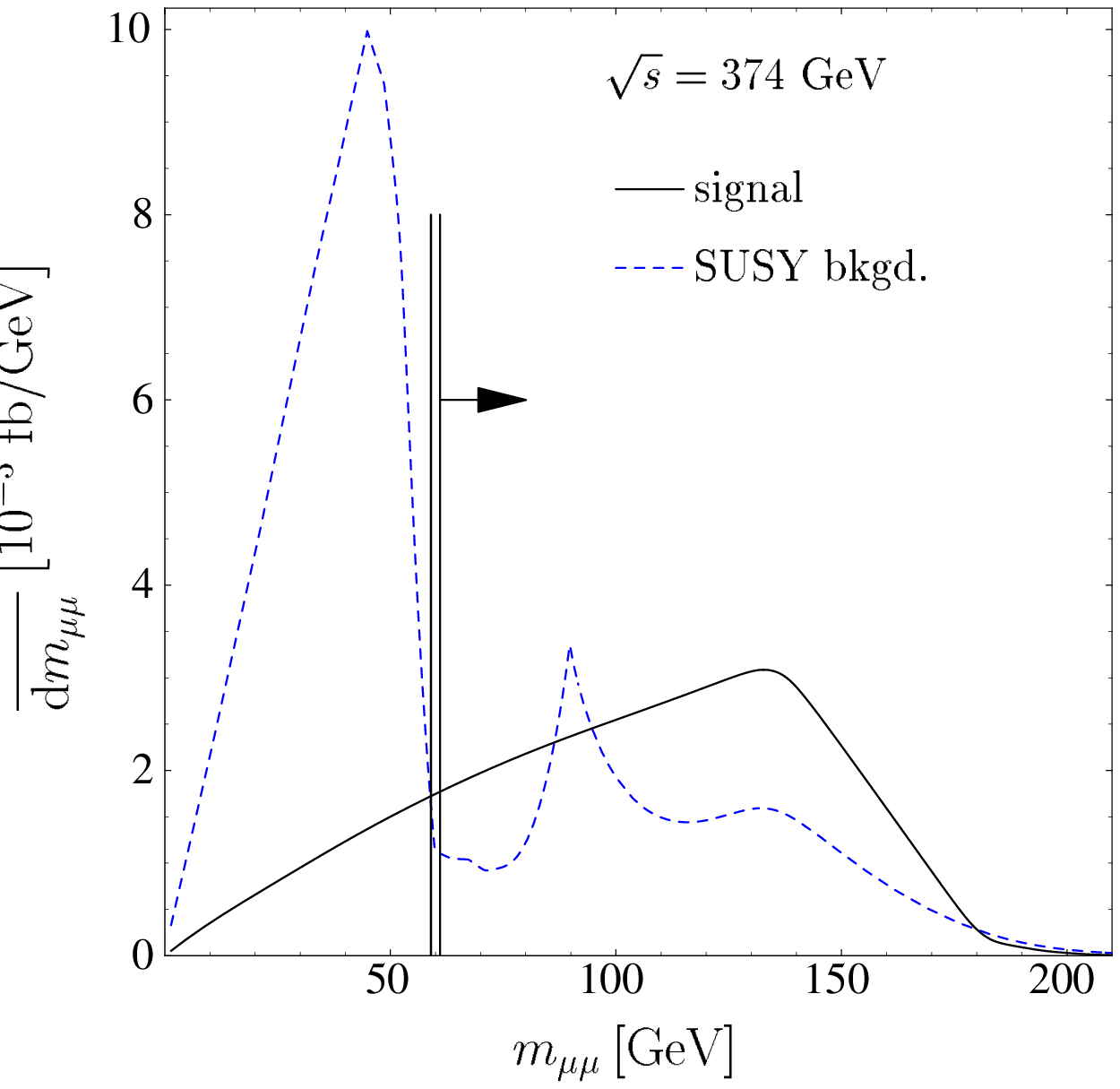,width=2.43in}
\hspace{0.7cm}
\epsfig{file=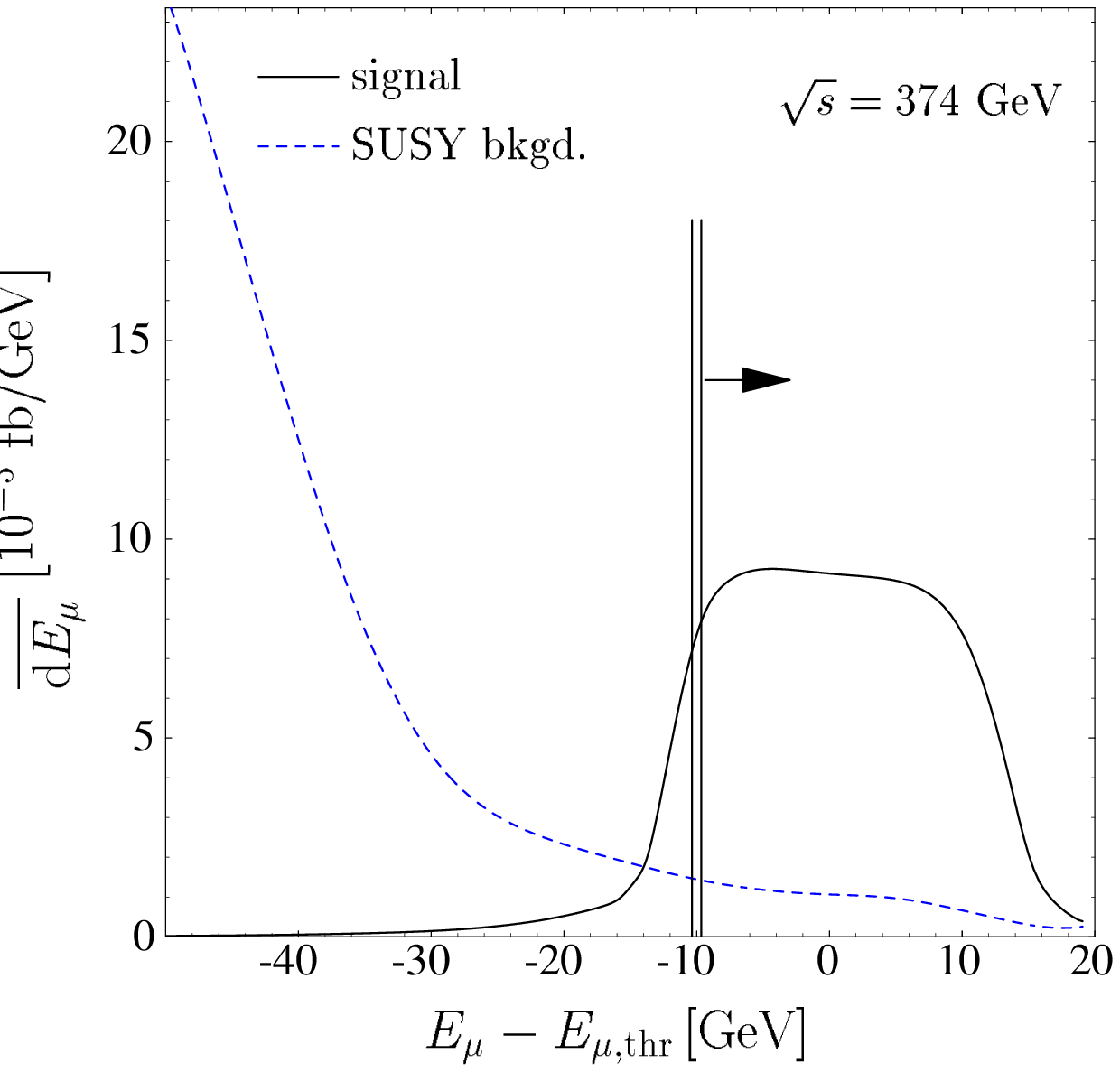,width=2.5in}
\end{center}
\caption{\it (a) The total cross-section and SUSY backgrounds to $e^+e^- \to \mu^+ \mu^- + \slsh{E}$, indicating the important nominal thresholds by arrows. 
  (b) The distributions in the muon-pair invariant mass (after the
  missing energy cut has been made) and in the muon energy, with possible
  cuts shown as a double vertical line.}
\label{fig:thres&cuts}
\end{figure}

Furthermore, a large number of SUSY backgrounds are involved, which had not been
all included so far.
Classes of background processes are 
catalogued in Tab.\hspace{.3ex}\ref{tabbkgd}.
\begin{table}
\begin{center}
\begin{tabular}{|l|ll|}
\hline
Source Process&Followed by&\\\hline\hline
$e^+e^- \to \tilde{\chi}_k^0 \tilde{\chi}_1^0$
&$\tilde{\chi}_k^0 \to \mu^+\mu^-\tilde{\chi}_1^0$&\\
$e^+e^- \to \tilde{\chi}_2^0 \tilde{\chi}_2^0$
&$\tilde{\chi}_2^0 \to \mu^+\mu^-\tilde{\chi}_1^0$ &
$\tilde{\chi}_2^0 \to \bar\nu \nu \tilde{\chi}_1^0$ \\
$e^+e^- \to \tilde{\chi}_1^+ \tilde{\chi}_1^-$
&$\tilde{\chi}_1^\pm \to \mu^{\pm} \nu_{\mu} \tilde{\chi}_1^0$&\\\hline
$e^+e^- \to ZZ$
&$Z \to \mu^+ \mu^- $ & $Z \to \tilde{\chi}_1^0 \tilde{\chi}_1^0$\\
$e^+e^- \to Z\,h_0/H_0$
&$Z \to \mu^+ \mu^- $ & 
$h_0/H_0 \to \tilde{\chi}_1^0 \tilde{\chi}_1^0$\\\hline
\end{tabular}
\caption{\it The  significant doubly-resonant SUSY background processes 
contributing to $e^+e^- \to \mu^+ \mu^- + \slsh{E}$ near the $\smu^+_{\mathrm{R}}\smu^-_{\mathrm{R}}$ production threshold. Singly-resonant 
and non-resonant SUSY backgrounds are also included in the analysis.}
\label{tabbkgd}
\end{center}
\end{table}
The most important cross-sections are shown in \fg\ref{fig:thres&cuts}(a). 
All these background  processes can be reduced to a controllable level by
applying cuts on the  muon energies and on the missing energy. The SM motivated
cuts of Ref.~$\!$\cite{Martyn} have only little effect on the SUSY induced
backgrounds and are not included in this analysis.

\vspace{2mm}
Background processes involve decay cascades which give rise to
threshold effects in the \underline{muon-pair invariant mass} and
increased \underline{missing energy} compared to the signal process. 
The large $\schi_2^0 \schi_1^0$ background can be reduced drastically 
by selecting events with invariant mass 
\be m_{\mu\mu} \, \gesim \, m_{\schi_2^0}-m_{\schi_1^0} \ee
which amounts to 60 GeV for the mass values of the reference point.
In addition, requiring the missing energy $\slsh{E}$ below the cut
\be \slsh{E} \lesim E_{\mathrm{cut}}=\sqrt{s} \left[ 1
       - \frac{(m_{\schi_2^0}^2-m_{\schi_1^0}^2)
         \left(s-m_{\schi_1^0}^2+m_{\schi_2^0}^2
       -\lambda^{1/2}(s,\mnsq{1},\mnsq{2})\right)}
         {4  \, m_{\schi_2^0}^2 \, s} \right]
\ee         
eliminates a large fraction of the background events from
$\schi_1^0\schi_2^0$ cascades to $\mu^+\mu^-\schi_1^0\schi_1^0$.  In
the scenario \RRtwo, events with missing energy below a fraction
$0.63$ of the c.m. energy are kept in the signal sample. Only a small
number of signal events are lost while the $\schi_1^0\schi_2^0$
background is reduced by a factor of fifty. After applying these two
cuts we are left with a small amount of higher cascade decays.
This is apparent from \fg\ref{fig:thres&cuts}(b/left) which shows the
invariant $\mu$-pair mass after the missing energy cut has already
been applied.

\vspace{2mm}
Near the threshold the observed \underline{muon energy} is sharp: 
\be E_{\mu} \approx (m_{\smu}^2-m_{\tilde{\chi}_1^0}^2)/2m_{\smu} \ee 
Alternatively to the $\slsh{E}$ cut, the signal to background ratio can also be 
greatly enhanced by selecting muons with energies in a band 
$\Delta E \approx 10$~GeV about the nominal threshold energies, cf. 
\fg\ref{fig:thres&cuts}(b/right).

\vspace{2mm}
The matrix elements are calculated using the computer algebra package
\mbox{\emph{FeynArts}~\cite{feynarts}.} Due to the large number of diagrams
involved, i.e. more than 300, it is convenient to perform the evaluation in
terms of helicity
amplitudes which are then evaluated using the Dirac spinor techniques
of Ref.~$\!$\cite{KS}.  The phase-space integration can be  performed by
multi-channel Monte-Carlo methods where appropriate phase-space
mappings may be used in order to obtain numerically stable results of
high accuracy~\cite{multich}. The Monte-Carlo error is reduced by
adaptive weight optimization \cite{weightopti}.  Initial state
radiation from emission of collinear and soft photons is included
using the structure-function \mbox{method \cite{ISR:struct}} up to second
order in the leading-log approximation and soft-photon \mbox{exponentiation
\cite{ISR:approx}}.  Beamstrahlung effects are also included using the
program {\it K$\iota${}$\rho${}$\kappa${}$\eta$}~\cite{circe}
for \TESLA accelerator design parameters.

\subsection{Results}

\begin{figure}[tb]
\centering\epsfig{file=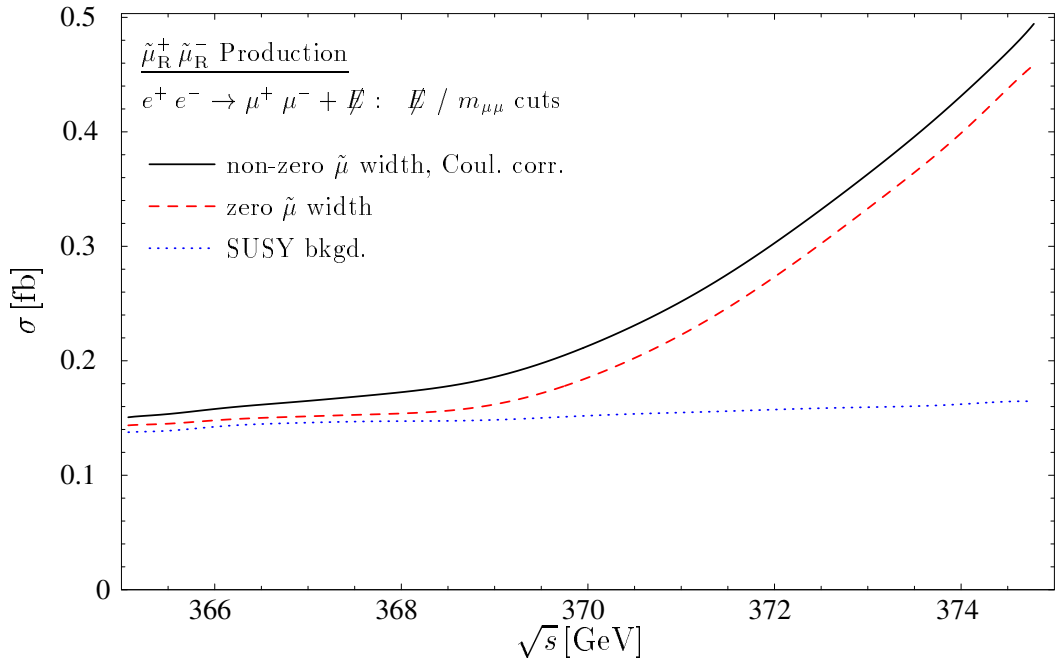,width=5.6in}\\
\caption{\it The excitation curves for the signal (with non-zero width and
Coulomb rescattering, compared with zero width) and the SUSY background after
the missing energy cut.
}
\label{fig:aftercuts}
\end{figure}

Including the SUSY backgrounds, the excitation curves, after
the beamstrahlung is switched on and the cuts are applied, are shown
in \fg\ref{fig:aftercuts}
for the missing-energy cut as a characteristic example (similar results follow
from the muon-energy cut).
As evident from the figure,
the background is smooth below the signal
curve in the threshold region.  The background can therefore be
extrapolated from below into the threshold region and subtracted
experimentally in a model-independent way since interference effects
are small.

\section{Selectron Excitation}

\subsection{Diagonal Selectron Pairs in {\boldmath $e^-e^-$} Collisions}

\begin{figure}[!b]
\begin{tabular}{p{7cm}p{7cm}}
 (a) \underline{$e^-e^- \to e^-e^- \schi_1^0 \schi_1^0$} \vspace{0.3cm}\newline
\phantom{(a} \epsfig{file=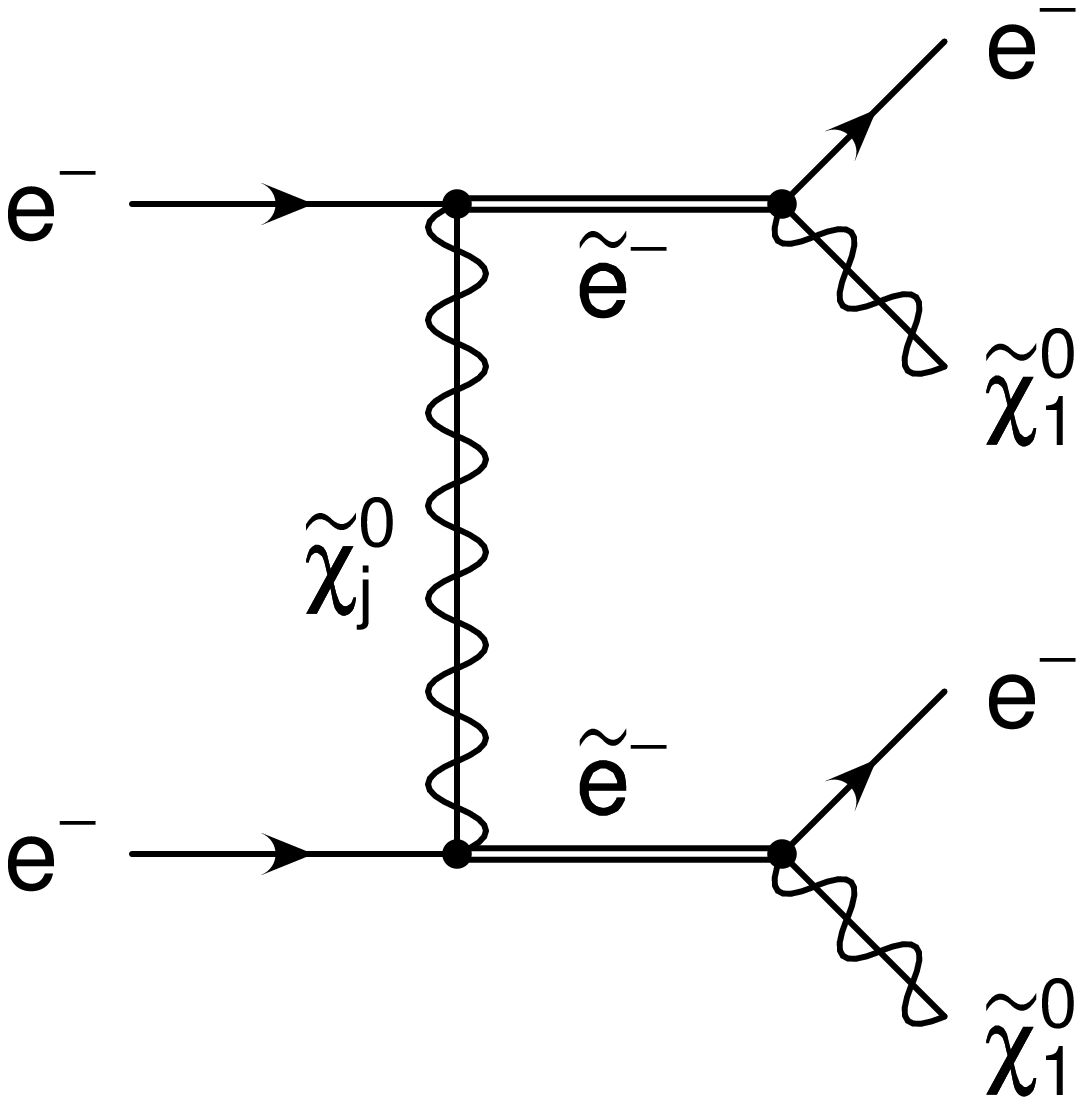,width=5cm} \vspace{0.5cm}\newline &
 (b) \underline{$e^-e^- \to \nu_e \nu_e \schi_1^- \schi_1^-$} \vspace{0.3cm}\newline
\phantom{(b} \epsfig{file=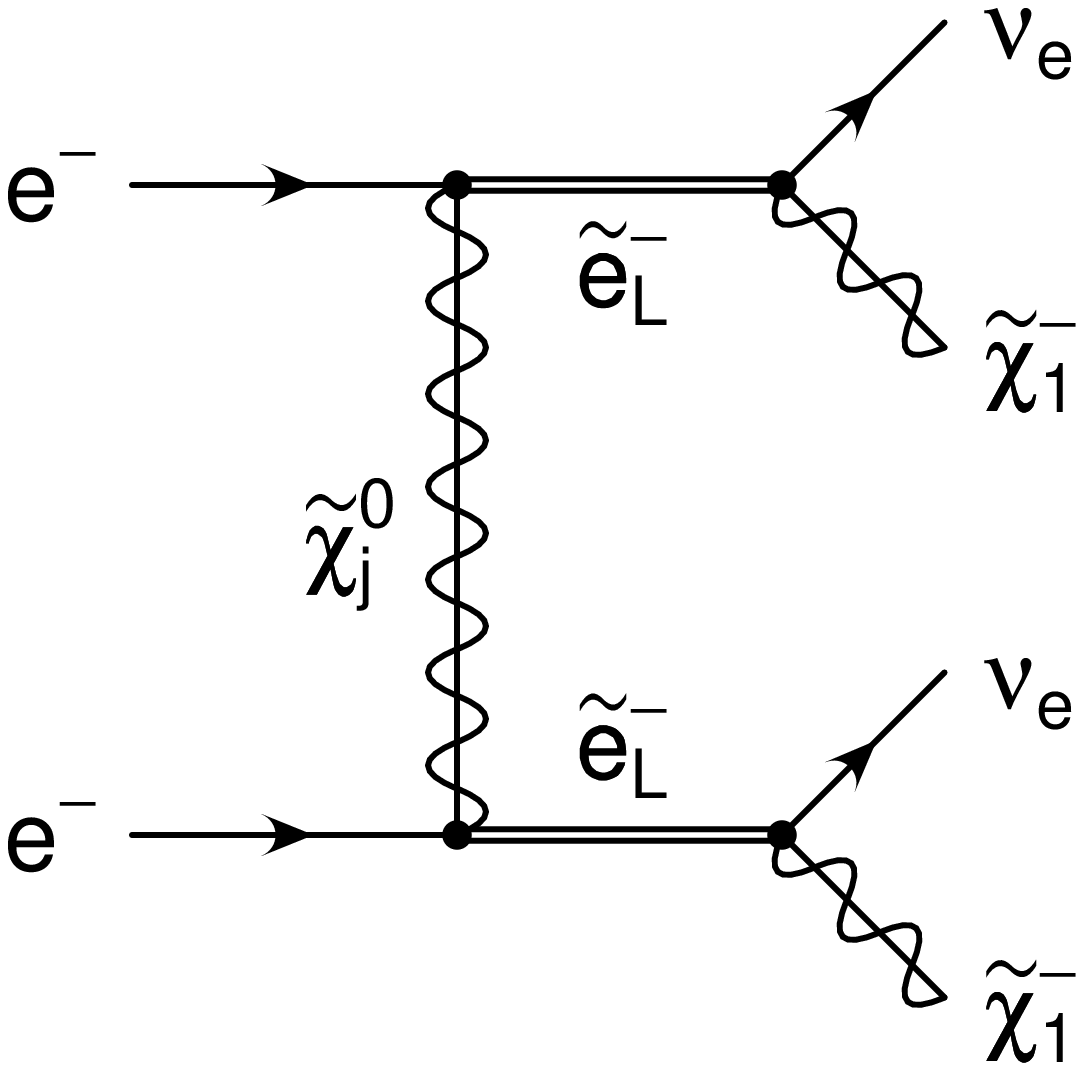,width=5cm}
\end{tabular}
\vspace{-1cm}
\caption{\it The doubly resonant contributions 
to the processes (a) \mbox{$e^-e^- \to e^-e^- \schi_1^0 \schi_1^0$} and (b)
\mbox{$e^-e^- \to \nu_e \nu_e \schi_1^- \schi_1^-$}.}
\label{fig:emem}
\end{figure}
Among the S-wave production processes for left- and right-chiral selectrons, the
$e^-e^-$ collisions are of particular interest. Even though mixed pairs in
$e^+ e^-$
collisions are also produced in S-waves, large SM and SUSY backgrounds
(in particular production of neutralino and chargino pairs as listed in
Tab.\hspace{.3ex}\ref{tabbkgd} and heavier states followed by cascade decays)
render these
channels less attractive. The analysis will therefore be presented in detail for
the channels $e^- e^- \to \seL^- \seL^-, \seR^- \seR^-$. The degree of
polarization for the electron beams will be assumed 80\%. In between the $\seR^-
\seR^-$ and $\seL^- \seL^-$ thresholds, the impurity of the electron helicity
states will also generate mixed $\seL^- \seR^-$ events which however are P-wave
suppressed.

Pairs of \underline{right-chiral selectrons $\seR^-$}
almost completely decay into the final state $e^- e^-
\schi^0_1 \schi^0_1$, resulting in the experimental signature of two electrons
plus missing energy.
The SM backgrounds predominantly arise from single $W$ and $Z$ production and
can be reduced with the help of beam polarization and cutting on invisibly
decaying $Z$ bosons \cite{COR:emem}. These cuts have little impact on the
signal. This study therefore focuses on the supersymmetric
backgrounds, which had not been included before.
The relevant SUSY backgrounds with the signature $e^- e^- + \slsh{E}$ do not
contain any pair production processes and turn out to be small over the full
energy scale under consideration.

Pairs of \underline{left-chiral selectrons $\seL^-$} can be selected by
choosing a unique final state. If kinematically accessible, the main decay
channel is the decay into charginos, $\seL^- \to \nu_e \, \schi^-_k$,
eq.~$\!$(\ref{deccha}), with a branching ratio of more than 50\%. It is convenient
to consider a leptonic decay channel of one of the charginos, $\schi^-_1 \to
l^- \, \bar{\nu}_l \, \schi^0_1$ with $l \neq e$, and a hadronic decay channel
of the second chargino, $\schi^-_1 \to q \, \bar{q}' \, \schi^0_1$. About 30\%
of the chargino pairs follow these combined decay modes. The resulting
signature $l \, q \, \bar{q}' + \slsh{E}$ is contaminated by only little SUSY
backgrounds\footnote{The SM backgrounds are kept small by this choice, too.}.

\subsection{Results}

\begin{figure}[tb]
\rule{0mm}{0mm}\hspace{-.26in}
\epsfig{file=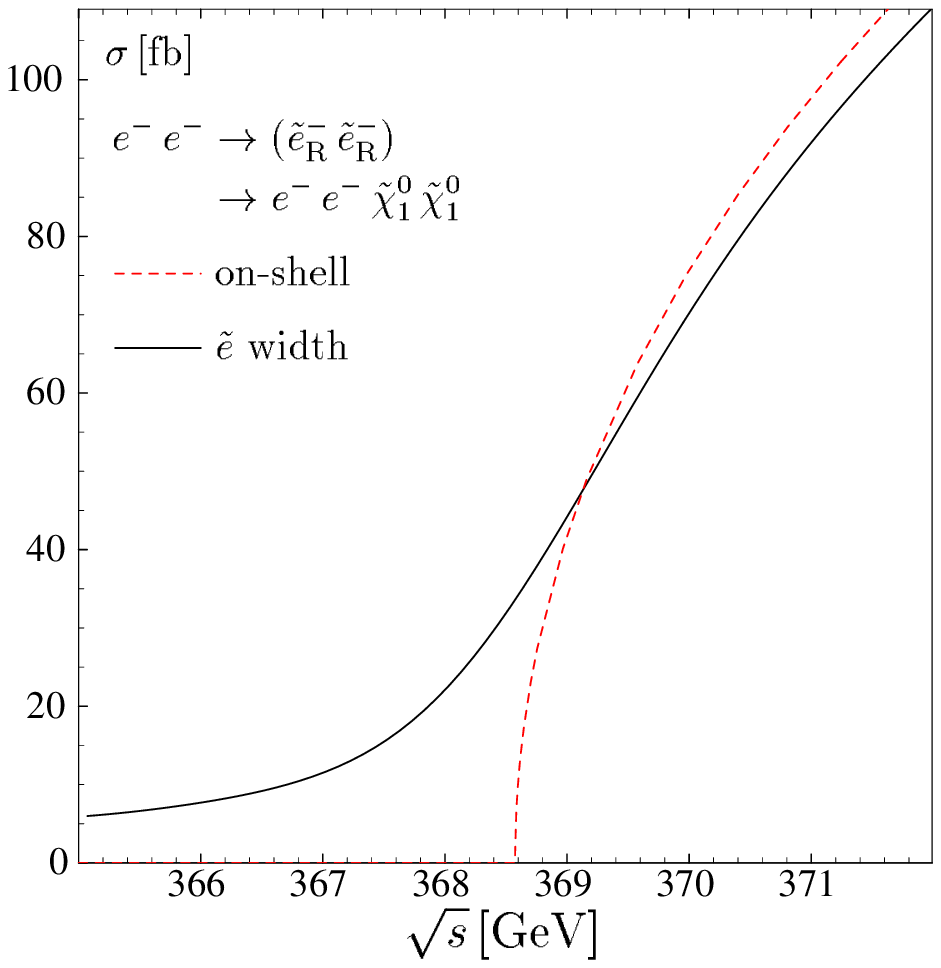,height=3.2in}
\hspace{-.27in}
\epsfig{file=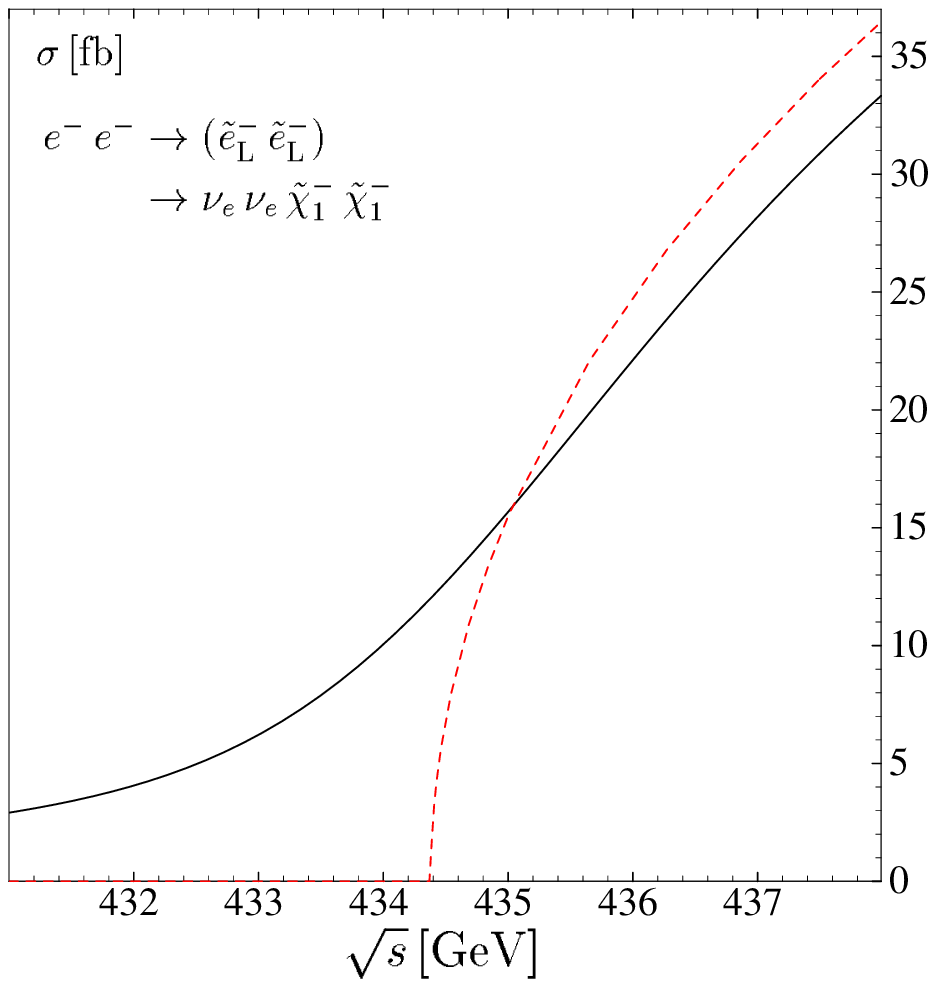,height=3.2in}
\caption{\it The effect of finite selectron widths on the
\mbox{$e^-e^- \to \se_{\rm R}^- \se_{\rm R}^- \to e^-e^-\schi_1^0\schi_1^0$}
and 
\mbox{$e^-e^- \to \se_{\rm L}^- \se_{\rm L}^- \to \nu_e \nu_e
\schi_1^-\schi_1^-$} cross-sections.}
\label{fig:sigee_at_thresh}
\end{figure}

The threshold cross-sections for $\eR^- \eR^- \to \seR^- \seR^-$ and $\eL^-
\eL^- \to \seL^- \seL^-$ are shown in \fg\ref{fig:sigee_at_thresh}
for zero and non-zero width, respectively.

Since the amplitudes corresponding to the diagrams in \fg\ref{fig:emem}
do not include any gauge boson exchanges, no specific
attention needs to be paid to gauge invariance.
These cross-sections, including
the non-zero width, may therefore be interpreted as ``signal cross-sections''.
However, for a systematic analysis, it is necessary to include all
contributions with the four fermion final states.
Coulombic photon rescattering effects are described by the modified Sommerfeld
correction with ${\mathcal C}_0 = 1$ in eq.~$\!$(\ref{offs}).

\begin{figure}[tb]
\rule{0mm}{0mm}\hspace{-.26in}
\epsfig{file=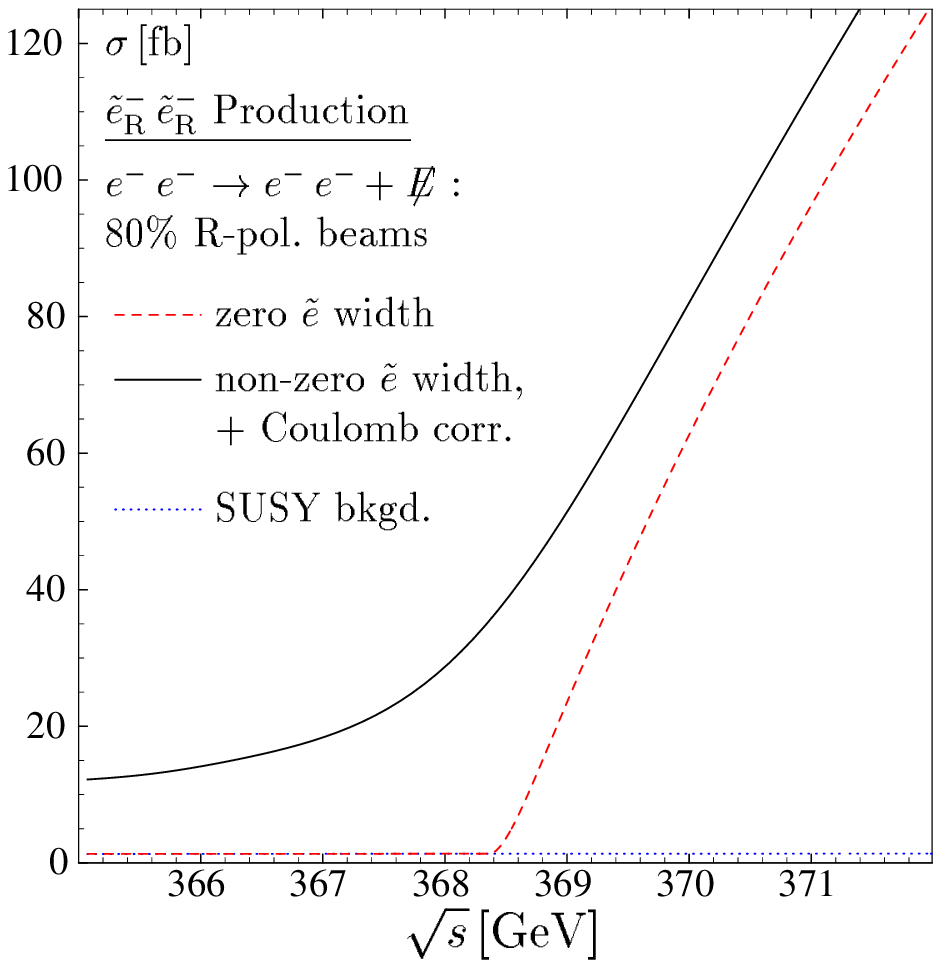,height=3.2in}
\hspace{-.27in}
\epsfig{file=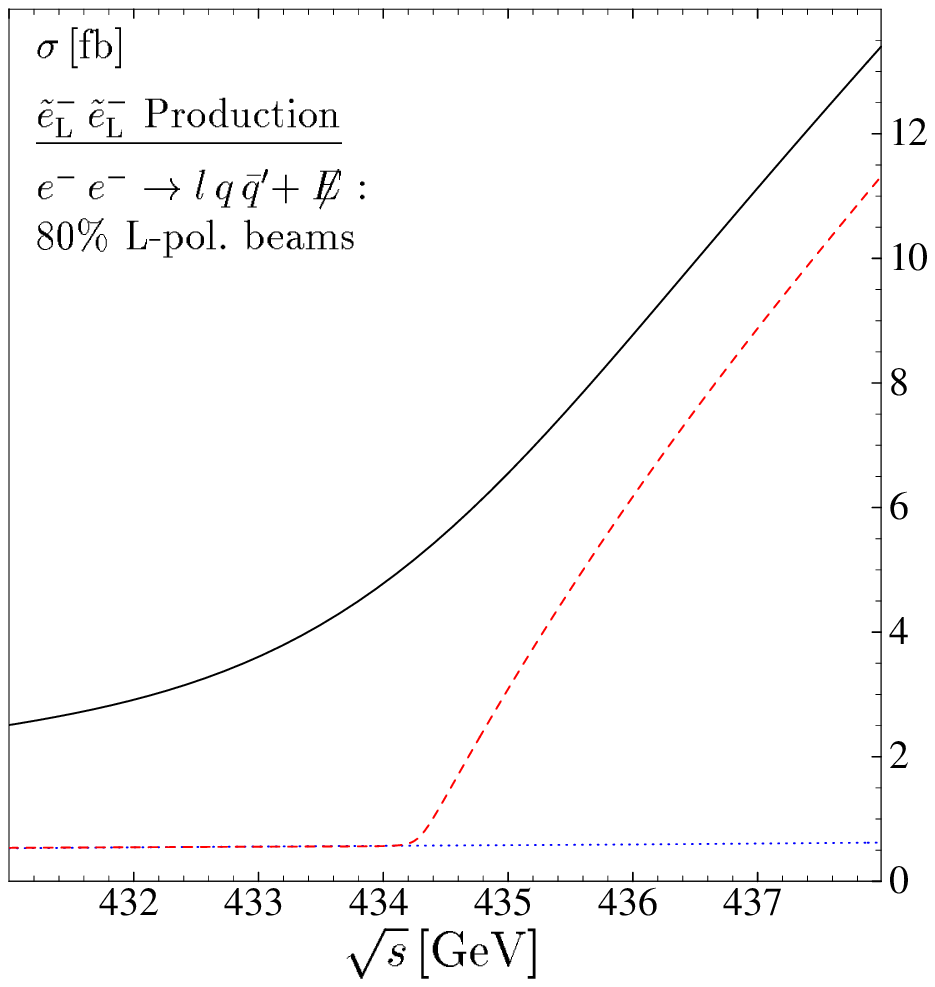,height=3.2in}
\caption{\it The excitation curves for the signal (with non-zero width and
Coulomb rescattering, compared with zero width) and the SUSY background 
near the $\seR\seR$ and $\seL\seL$ thresholds.}
\label{fig:ememthr}
\end{figure}

The final results are presented in \fg\ref{fig:ememthr}.
The figures prove that contributions from SUSY backgrounds are small and do not
require additional cuts. They exhibit
a flat energy dependence, thereby easily allowing
their experimental subtraction
from the excitation curves. The effect of Coulomb rescattering is
more pronounced for S-wave threshold production in contrast to the P-wave 
smuon case.

\section{Summary}

The present report describes a first theoretical step into the area
of high-precision analyses in supersymmetric theories. We have concentrated on
non-zero width and Coulomb rescattering effects in the threshold production of
smuon and selectron pairs. Moreover, SUSY backgrounds to SUSY signals are
systematically analyzed.

For smuon pair production, $e^+e^- \to \smR^+ \smR^-, \smL^+ \smL^-$, proper
attention must be paid to the gauge invariance of the amplitudes when non-zero
widths are included. These effects are of the same size as the expected
experimental resolution. In contrast to P-wave smuon production, which is
$\propto \beta^3$ near threshold, selectron pairs
can be generated in S-waves with a steep rise $\propto \beta$ of the threshold
cross-section. This advantage can be exploited in $e^+e^-$ and $e^-e^-$
collisions. However, SUSY and SM backgrounds in $e^-e^-$ collisions are
significantly smaller, so that $\eR^- \eR^- \to \seR^- \seR^-$ and $\eL^-
\eL^- \to \seL^- \seL^-$ are the preferred production processes for right- and
left-chiral selectrons $\seR$ and $\seL$.

It is evident that many additional detailed calculations of
genuine higher order effects must follow to reach final theoretical accuracies
at a level of a per-mille as required by the expected accuracy of $e^+e^-$
linear colliders.

\vspace{3ex}
\noindent
{\bf Note:} Shortly before completing the manuscript we learned about a similar
analysis of non-zero width effects in the threshold production of
right-chiral selectrons at electron colliders by
J.~L.~Feng and M.~E.~Peskin [hep-ph/0105100].

\section*{Acknowledgments}

We thank A.~von~Manteuffel and H.~U.~Martyn for useful discussions.
Correspondence with V.~Fadin and V.~A.~Khoze is gratefully acknowledged.

\clearpage
\newpage

\section*{Appendix}

\noindent Reference scenario \RRtwo with $\tan\beta=30$ $\;(l = e, \mu)$:

\renewcommand{\arraystretch}{1.3}

\vspace{2ex}
\noindent
\parbox[t]{4.45cm}{
{\bf Neutralino masses:} \\[0.165cm]
\begin{tabular}{|c|r@{.}l|}
\hline
\hspace{2.5em} & \multicolumn{2}{c|}{$m_{\tilde{\chi}^0_i}$ [GeV]}\\
\hline
$\tilde{\chi}^0_1$ & 74&81 $\;$ \\
$\tilde{\chi}^0_2$ & $\;$ 133&04 \\
$\tilde{\chi}^0_3$ & 272&81 \\
$\tilde{\chi}^0_4$ & 292&96 \\
\hline 
\multicolumn{3}{l}{$\!\!\!$\bf Chargino masses:}\\
\hline
\hspace{2.5em} & \multicolumn{2}{c|}{$m_{\tilde{\chi}^\pm_i}$ [GeV]} \\
\hline
$\tilde{\chi}^\pm_1$ & 132&35 $\;$ \\
$\tilde{\chi}^\pm_2$ & $\;$ 294&84 \\
\hline
\end{tabular}
}
\parbox[t]{5.8cm}{
{\bf Right-chiral slepton {\boldmath $\ts{\tilde{l}}{R}$}:}\\[0.5ex]
\begin{tabular}{|r@{}l|r@{.}l@{ }l|}
\hline
\multicolumn{2}{|c|}{Mass $m_{\tilde{l}}$} & 184&29 &$\!\!\!$GeV \\
\multicolumn{2}{|c|}{Width $\Gamma_{\tilde{l}}$} & 0&62 &$\!\!\!$GeV \\
\hline
\multicolumn{5}{c}{\hspace{0.2cm}}\\
\hline
\multicolumn{5}{|c|}{Branching Ratios}\\
\hline
$\ts{\tilde{l}}{R}^- $&$\to l^- \, \tilde{\chi}^0_1$ & 0&991& \\
&$ \to l^- \, \tilde{\chi}^0_{2}$ & 0&008& \\
&$ \to l^- \, \tilde{\chi}^0_{3,4}$ & \multicolumn{2}{c}{---}& \\
\hline
\end{tabular}
}
\parbox[t]{5cm}{
{\bf Left-chiral slepton {\boldmath $\ts{\tilde{l}}{L}$}:}\\[0.5ex]
\begin{tabular}{|r@{}l|r@{.}l@{ }l|}
\hline
\multicolumn{2}{|c|}{Mass $m_{\tilde{l}}$} & 217&19 &$\!\!\!$GeV \\
\multicolumn{2}{|c|}{Width $\Gamma_{\tilde{l}}$} & 1&03 &$\!\!\!$GeV \\
\hline
\multicolumn{5}{c}{\hspace{0.2cm}}\\
\hline
\multicolumn{5}{|c|}{Branching Ratios}\\
\hline
$\ts{\tilde{l}}{L}^- $&$\to l^- \, \tilde{\chi}^0_1$ & 0&144 &\\
&$\to l^- \, \tilde{\chi}^0_{2}$ & 0&338 &\\
&$\to l^- \, \tilde{\chi}^0_{3,4}$ & \multicolumn{2}{c}{---}&\\
&$\to \nu_l \, \tilde{\chi}^-_{1}$ & 0&517 &\\
&$\to \nu_l \, \tilde{\chi}^-_{2}$ & \multicolumn{2}{c}{---}&\\
\hline
\end{tabular}
}

\end{document}